\documentclass[11pt,twoside]{article}

\usepackage{fancyhdr}
\usepackage[colorlinks,citecolor=blue,urlcolor=blue,linkcolor=blue,bookmarks=false]{hyperref}
\usepackage{amsfonts,epsfig,graphicx}
\usepackage{afterpage}
\usepackage{comment}
\usepackage{amsmath,amssymb,amsthm} 
\usepackage{fullpage}
\usepackage[T1]{fontenc} 
\usepackage{epsf} 
\usepackage{graphics} 
\usepackage{enumitem} 
\usepackage{amsfonts,amsmath}
\usepackage[sort]{natbib} 
\usepackage{psfrag,xspace}
\usepackage{color,etoolbox}
\usepackage{subcaption}
\usepackage{float}

\def\indist{\rightsquigarrow}
\def\ind{\perp\!\!\!\perp}

\usepackage{amssymb,mathtools}

\DeclarePairedDelimiter{\Norm}{\lVert}{\rVert}

\newcommand{\var}{\text{var}}

\newcommand{\Pb}{\mathbb{P}}
\newcommand{\Pn}{\mathbb{P}_n}

\newcommand{\E}{\mathbb{E}}
\newcommand{\R}{\mathbb{R}}

\newcommand{\pihat}{\widehat{\pi}}
\newcommand{\muhat}{\widehat{\mu}}

\newcommand{\psihat}{\widehat{\psi}}

\DeclareSymbolFont{bbold}{U}{bbold}{m}{n}
\DeclareSymbolFontAlphabet{\mathbbold}{bbold}
\newcommand{\one}{\mathbbold{1}}
\newtheorem{theorem}{Theorem}

\newtheorem{algorithm}{Algorithm}
\theoremstyle{definition}

\theoremstyle{remark}
\newtheorem{assumption}{Assumption}

\newtheorem{remark}{Remark}

\begin{document}

\def\spacingset#1{\renewcommand{\baselinestretch}{#1}\small\normalsize} \spacingset{1}

\raggedbottom
\allowdisplaybreaks

%%%%%%%%%%%%%%%%%%%%%%%%%%%%%%%%%%%%%%%%%%%

\title{\vspace*{-.4in} {Incremental causal effects: an introduction and review}}
\author{\\ Matteo Bonvini,$^*$ Alec McClean,\footnote{MB and AM contributed equally.} \ Zach Branson, Edward H. Kennedy \\ \\
Department of Statistics \& Data Science \\
Carnegie Mellon University \\ \\
\texttt{\{mbonvini, zach, edward\} @stat.cmu.edu, amcclean@andrew.cmu.edu} \\
\date{}}

\maketitle
\thispagestyle{empty}

%%%%%%%%%%%%%%%%%%%%%%%%%
%%% Abstract

\begin{abstract}
In this chapter, we review the class of causal effects based on incremental propensity scores interventions proposed by \cite{kennedy2019nonparametric}. The aim of incremental propensity score interventions is to estimate the effect of increasing or decreasing subjects' odds of receiving treatment; this differs from the average treatment effect, where the aim is to estimate the effect of everyone deterministically receiving versus not receiving treatment. We first present incremental causal effects for the case when there is a single binary treatment, such that it can be compared to average treatment effects and thus shed light on key concepts. In particular, a benefit of incremental effects is that positivity---a common assumption in causal inference---is not needed to identify causal effects. Then we discuss the more general case where treatment is measured at multiple time points, where positivity is more likely to be violated and thus incremental effects can be especially useful. Throughout, we motivate incremental effects with real-world applications, present nonparametric estimators for these effects, and discuss their efficiency properties, while also briefly reviewing the role of influence functions in functional estimation. Finally, we show how to interpret and analyze results using these estimators in practice, and discuss extensions and future directions.
\end{abstract}


\begin{comment}
TODO:
\begin{itemize}
    \item Uniform notation (use biometrika)
    \item Add references;
    \item uniform inference one-time-pt case (should we include the multiplier bootstrap?);
    \item add figures / expand section that describes intuition behind incremental effects;
    \item expand MSM vs incremental effects in multiple time pt case;
    \item simulations or data analysis?
    \item alternative interventions?
\end{itemize}
\end{comment}

%%%%%%%%%%%%%%%%%%%%%%%%%
%%% Introduction

\section{Introduction} \label{sec:intro}

Understanding the effect of a variable $A$, the treatment, on another variable $Y$, the outcome, involves measuring how the distribution of $Y$ changes when the distribution of $A$ is manipulated in some way. By ``manipulating a distribution,'' we mean that we imagine a world where we can change the distribution of $A$ to some other distribution of our choice, which we refer to as the ``intervention distribution.'' The intervention distribution defines the causal effect of $A$ on $Y$, i.e., the causal estimand. For instance, suppose $A$ is binary and is completely randomized with some probability $c_0$; one may then ask how the average outcome $Y$ would change if the randomization probability were set to some other constant $c_1$.

\medskip

In order to answer questions about causal effects, we adopt the potential outcomes framework (\cite{rubin1974estimating}). We suppose that each subject in the population is linked to a number of ``potential outcomes,'' denoted by $Y^a$, that equal the outcome $Y$ that would have been observed if the subject had received treatment $A = a$. In practice, only one potential outcome is observed for each subject, because one cannot go back in time and assign a different treatment value to the same subject. As a result, we must make several assumptions to identify and estimate causal effects. One common assumption is ``Positivity,'' which says that each subject in the population must have a non-zero chance of receiving each treatment level. Positivity is necessary to identify the average treatment effect, which is the most common causal estimand in the literature; however, Positivity may be very unrealistic in practice, particularly in many time-point analyses where the number of possible treatment regimes scales exponentially with the number of time-points. 

\medskip

In this chapter, we introduce and review the class of ``incremental causal effects'' discussed in \cite{kennedy2019nonparametric}. Incremental causal effects are based on an intervention distribution that shifts the odds of receiving a binary treatment by a user-specified amount. Crucially, incremental causal effects are well-defined and can be efficiently estimated even when Positivity is violated. Furthermore, they represent an intuitive way to summarize the effect of the treatment on the outcome, even in high-dimensional, time-varying settings. In Section \ref{sec:prelim} we first review how to identify and estimate the average treatment effect, as well as review the notion of static interventions, dynamic interventions, and stochastic interventions. A limitation of the average treatment effect is that it only considers a static and deterministic intervention, where all subjects are either assigned to treatment or control, which may be improbable if Positivity is violated or nearly violated. On the other hand, dynamic and stochastic interventions allow the treatment to vary across subjects, thereby lessening dependence on Positivity. Incremental causal effects assume a particular stochastic intervention, which we discuss in depth for binary treatments in Section \ref{sec:incremental} before discussing time-varying treatments in Section \ref{sec:time}. We then demonstrate how to use incremental effects in practice in Section \ref{sec:example}. We conclude with a discussion of extensions and future directions for incremental causal effects in Section \ref{sec:extensions}.

%%%%%%%%%%%%%%%%%%%%%%%%%
%%% Preliminaries

\section{Preliminaries: Potential Outcomes, the Average Treatment Effect, and Types of Interventions} \label{sec:prelim}

In this section, to introduce and focus ideas, we consider studies with a single binary exposure of interest. Longitudinal studies with time-varying exposures are presented in Section \ref{sec:time}. Here, for each of $n$ subjects, we observe a binary treatment $A$, a vector of covariates $X$, and an outcome $Y$, each indexed by $i = 1,\dots,n$. The goal is to estimate the causal effect of the treatment $A$ on the outcome $Y$, with the complication that subjects self-select into treatment in a way that may depend on the covariates $X$. A fundamental quantity in this setting is the probability of receiving treatment $\pi(X) \equiv \Pb(A = 1 \mid X)$, known as the propensity score. We also let $\one(\mathcal{E})$ be a binary indicator for whether event $\mathcal{E}$ occurs.  

\medskip

When the treatment is binary and there is no interference, each subject has only two potential outcomes, $Y^1$ and $Y^0$. As a running example, we will consider estimating the effect of behavioral health services on probationers' likelihood of re-arrest.  Recidivism is a colossal societal issue in the United States.  Millions of people are on probation in the U.S. and around half of the people who leave prison are re-arrested within one year \citep{bjs2021keystatistics, bjs2018recidivism}.  Further, outcomes may be even worse for people with mental illness or substance use disorders \citep{steadman2009prevalence, skeem2011correctional}. Because of this, some researchers have posited that attending behavioral health services (e.g., talk therapy) during probation may help probationers avoid criminal behavior and being re-arrested. In this example, the covariate information $X$ for each probationer includes age, gender, race, etc., the treatment $A$ denotes whether a probationer attended a behavioral health service over, say, the first six months of probation, and the outcome $Y$ denotes whether they were re-arrested within a year. The goal in this example is to assess if attending services lowers one's chance of being re-arrested. In Section \ref{sec:time} we  consider the more complex case where the covariates $X$, treatment $A$, and outcome $Y$ can be time-varying.

\medskip

First we will consider estimating the average treatment effect for this example. The average treatment effect implies a so-called \emph{static deterministic} intervention that assumes two extreme counterfactual scenarios, where either every subject receives treatment ($A = 1$) or every subject receives control ($A = 0$), which may not be realistic for many applications (including the above example from criminology). Then we will consider estimating incremental effects, which instead posit a stochastic dynamic intervention that interpolates counterfactual scenarios in between the ``everyone receives treatment'' and ``everyone receives control'' scenarios.

\subsection{Average treatment effects} \label{subsec:ates}

We first review the \emph{average treatment effect} (ATE), because it is by far the most common causal estimand of interest. The ATE is defined as:
\begin{align}
\text{ATE} \equiv \E (Y^1 - Y^0). \label{eqn:ate}
\end{align}
The ATE compares two quantities: $\mathbb{E}(Y^1)$, the mean outcome when all subjects receive treatment; and $\mathbb{E}(Y^0)$, the mean outcome when all subjects receive control. Within our example, these are the recidivism rate when everyone attends behavioral health services and the recidivism rate when no one attends behavioral health services, respectively.

The fundamental problem of causal inference (\cite{holland1986statistics}) is that each subject receives either treatment or control---never both---and thus we only observe one of the two potential outcomes for each subject. In other words, the difference $Y^1 - Y^0$ is only partially observed for each subject. Thus, assumptions must be made in order to identify and thereby estimate the ATE. It is common to make the following three assumptions to estimate the ATE:

\begin{assumption} \label{asm:cons} (Consistency). $Y = Y^a$ if $A = a$. \end{assumption}
\begin{assumption} \label{asm:exch} (Exchangeability). $A \ind Y^a \mid X$ for $a = 1$ and $a = 0$. \end{assumption}
\begin{assumption} \label{asm:pos} (Positivity). There exists $\epsilon > 0$ such that $\Pb( \epsilon \leq \pi(X) \leq 1 - \epsilon) = 1$. \end{assumption}

\medskip

Consistency says that if an individual takes treatment $a$, we observe their potential outcome under that treatment; Consistency allows us to write the observed outcomes as $Y = AY^1 + (1-A)Y^0$ and would be violated if, for example, there were interference between subjects, such that one subject's treatment affected another subject's outcome. Meanwhile, Exchangeability says that treatment is effectively randomized within covariate strata, in the sense that treatment is independent of subjects' potential outcomes---as in a randomized experiment---after conditioning on covariates. This assumption is also called ``no unmeasured confounding,'' which means that there are no variables beyond $X$ that induce dependence between treatment and the potential outcomes. Finally, Positivity says that the propensity score is bounded away from zero and one for all subjects. With these three assumptions, we have
$$
\E(Y^{a}) = \E\{\E(Y^{a} \mid X)\} = \E\{ \E(Y^a \mid X, A = a) \} = \E\{ \E(Y \mid X, A = a) \} = \E\left\{\frac{Y\one(A = a)}{\pi(X)}\right\},
$$
i.e., the mean outcome if all were assigned $A=a$ reduces to the regression function $\E(Y \mid X,A=a)$, averaged over the distribution of $X$. The average treatment effect is thus $\text{ATE} = \E\{ \E(Y \mid X, A = 1) - \E(Y \mid X, A = 0) \}$. 

\subsection{When positivity is violated}

We will give special attention to Positivity, for several reasons; it is often violated in practice (\cite{westreich2010invited,petersen2012diagnosing}), and it is also the only of the three causal assumptions that is in principle testable. It is important to remember that the ATE characterizes what would happen in two extreme counterfactual scenarios: one counterfactual world where  every subject receives treatment, versus another where every subject receives control. Although we never observe either of these two extreme scenarios, they nonetheless must be plausible in order for the ATE to be a sensible estimand to target. In other words, it must be plausible that every subject could receive treatment or control---i.e., Positivity must hold. Positivity would be violated if there are subjects for whom some treatment level is essentially impossible to receive. For example, in our running criminology example, some probationers might be required to attend behavioral health services as a condition of their probation, and thus their probability of attending treatment is one; conversely some probationers may be extremely unlikely to go to treatment. For both these groups, positivity would not hold. In this case, the ATE in (\ref{eqn:ate}) may not be a sensible estimand to target, since there is essentially zero chance of observing $Y^0$ or $Y^1$ for some subjects.

\medskip

In some settings researchers may know \textit{a priori} that Positivity does not hold for a particular application, while in others they may be concerned that Positivity does not hold after conducting preliminary exploratory data analysis (e.g., if estimated propensity scores for some subjects are close to zero or one). Furthermore, even if Positivity technically holds in a population, near-positivity violations in a sample can have disastrous effects on estimators for the ATE and related estimands. For example, it is well-known that inverse propensity score weighting and doubly robust estimators can suffer from high variance when propensity scores are extreme (\cite{kish1992weighting,busso2014new}). This does not mean that these estimators are unsatisfactory---when propensity scores are close to zero or one, at least one of the two extreme scenarios the ATE considers (``everyone receives treatment'') or (``everyone receives control'') is unlikely, and high variance is thus an appropriate signal of the uncertainty we have about those scenarios. Simple regression-based estimators may mask this high variance, but really only through extrapolation, which merely trades high variance for high bias. In fact, it can be shown (\cite{hahn1998role}) that, under exchangeability, no regular estimator of the ATE can have asymptotic variance smaller than
\begin{align*}
\E\left(\frac{\var(Y^1 \mid X)}{\pi(X)} + \frac{\var(Y^0 \mid X)}{1 - \pi(X)} + \left[\E(Y^1 - Y^0 \mid X) - \E\{\E(Y^1 - Y^0 \mid X)\} \right]^2\right).
\end{align*}
This is called the semiparemtric efficiency bound, which is the equivalent notion to the Cramer-Rao lower bound in parametric models. See \cite{newey1990semiparametric} for a review on this topic and the precise definition of regular estimators. Because the efficiency bound involves the reciprocal of the propensity score, it is clear that near-positivity violations have a deleterious effect on the precision with which the ATE can be estimated. 

\medskip

In the presence of positivity violations where there are substantial limits on how well one can estimate the ATE in \eqref{eqn:ate}, an alternative option is to target different estimands. For example, matching methods restrict analyses to a subsample that exhibits covariate balance between treatment and control, and thus positivity can be more plausible for that subsample (\cite{ho2007matching,stuart2010matching}). In this case, the targeted estimand would be the ATE \textit{for the matched subsample} instead of the ATE in the entire population. Similarly, propensity score trimming aims to remove subjects for whom propensity scores are outside the interval $[\epsilon, 1 - \epsilon]$, for some user-specified $\epsilon > 0$ (\cite{crump2009dealing}). If the true propensity scores were known, this trimming would ensure positivity holds by design, and the targeted estimand is the ATE \textit{for the trimmed subsample} instead of the ATE in (\ref{eqn:ate}). A benefit of these approaches is that standard ATE estimators can still be used, simply within a subsample. However, complications arise because the estimand then depends on the sample: The subsample ATE is only defined after a matching algorithm is chosen or propensity scores are estimated. Thus, interpretability may be a concern, because the causal effect is only defined for the sample at-hand instead of the broader population of interest. A way to overcome this shortcoming is to define the estimand based on the trimmed true propensity scores, i.e. $\E\{Y^1 - Y^0 \mid \pi(X) \in [\epsilon, 1 - \epsilon]\}$. However, estimating the quantity $\one\{\pi(X) \in [\epsilon, 1 - \epsilon]\}$ is challenging in flexible, nonparametric models because it is a non-smooth transformation of the data-generating distribution; as such, $\sqrt{n}$-consistent estimators do not generally exist without imposing further assumptions. 

\medskip

The estimands discussed so far---the ATE in (\ref{eqn:ate}) and subsample ATEs implied by matching or propensity score trimming---all represent \textit{static interventions}, where the treatment $A$ is set to fixed values across a population of interest. In what follows, we will consider alternative estimands that researchers can target in the face of positivity violations. These estimands concern effects of \emph{dynamic} (non-static) interventions, where, in counterfactual worlds, the treatment can be allowed to vary across subjects, instead of being set to fixed values. Indeed, the static interventions discussed so far---where every subject receives treatment or every subject receives control---may be impossible to implement for many applications, whereas non-static interventions may be closer to what we would expect to be possible in practice. 

\subsection{Dynamic interventions}

\emph{Dynamic interventions} allow the intervention to depend on subjects' covariate information; thus, the intervention is allowed to vary across subjects, depending on their covariate values, albeit still in a deterministic way. Dynamic interventions often arise in medical situations, where treatment is only given to people with severe symptoms, and thus Positivity only holds for a subset of the population. In this sense, dynamic interventions are quite similar to the interventions implied by matching or propensity score trimming, discussed at the end of the previous subsection. In particular, matching or propensity score trimming considers a \textit{static} intervention that only occurs within a subpopulation (which is often defined in terms of covariates or propensity scores), whereas dynamic interventions consider an intervention that occurs across the whole population, but the treatment for each subject will depend on their covariate values.

\medskip

As an example, let's say we want to measure the effect of providing behavioral health services only for probationers for whom Positivity holds, i.e., for probationers for whom $\mathbb{P}(A = 1 \mid X) \in [\epsilon, 1 - \epsilon]$. In other words, we would like to answer the question, "What is the causal effect of providing behavioral health services to probationers who can plausibly choose whether or not to attend services?" We could address this question by considering the following dynamic intervention:
\begin{align*}
    d_a(X) = \begin{cases}
    a &\mbox{ if } \mathbb{P}(A = 1 | X) \in [\epsilon, 1 - \epsilon] \\
    1 &\mbox{ if } \mathbb{P}(A = 1 | X) > 1 - \epsilon \\
    0 &\mbox{ if } \mathbb{P}(A = 1 | X) < \epsilon
    \end{cases}, \hspace{0.2in} a \in \{0, 1\}
\end{align*}
A similar dynamic intervention was discussed in \cite{moore2012causal}. Under this intervention, we set subjects to treatment $a$ when Positivity holds; otherwise, treatment is fixed at $a = 1$ for subjects that will almost certainly receive treatment and $a = 0$ for subjects that will almost certainly receive control. In this case, the causal estimand is $\mathbb{E} \left\{ Y^{d_1(X)} - Y^{d_0(X)} \right\}$, where $\mathbb{E} \left\{ Y^{d_a(X)} \right\}$ denotes the average outcome when $A$ is set according to $d_a(X)$ across subjects. Note that $\mathbb{E} \left\{ Y^{d_1(X)} \right\}$ differs from $\mathbb{E}(Y^1)$ in the ATE (\ref{eqn:ate}), to the extent that there are subjects for whom $\Pb(A = 1 \mid X) < \epsilon$, in which case some subjects receive control when estimating $\mathbb{E} \left\{ Y^{d_1(X)} \right\}$ but not $\mathbb{E} (Y^1)$. In this case, $\mathbb{E} \left\{ Y^{d_1(X)} - Y^{d_0(X)} \right\}$ is equivalent to the ATE within a propensity score trimmed subsample, but the hypothetical intervention is employed across the entire population.

\medskip

More generally, a dynamic intervention implies causal effects on the subpopulation whose treatment is affected by that dynamic intervention. Because practitioners may realistically only be able to intervene on a subset of the population, causal effects implied by dynamic interventions may be relevant for many applications. That said, dynamic interventions assume that we can deterministically assign treatment or control to all subjects as a function of their covariates. In reality, treatment assignment may still be stochastic when an intervention is employed. In this case, the intervention is considered a stochastic intervention.

\subsection{Stochastic interventions}

A \textit{stochastic intervention} assigns treatment randomly based on a probability distribution. As a simple example, consider a Bernoulli trial, where we assign each subject to treatment with probability $p$, and otherwise we assign them to control. In this case, the probability distribution of treatment for each subject is $Q(a) = p^a(1-p)^{1-a}$ for $a \in \{0,1\}$, and we might consider the causal effect of changing the parameter $p$. Under this intervention, although we control subjects' probability of receiving treatment, the actual realization of treatment is stochastic, unlike the static and dynamic interventions we've discussed so far. Stochastic interventions can be considered generalizations of static and dynamic interventions; note that the static intervention implied by the ATE in (\ref{eqn:ate}) is a special case of the Bernoulli trial example, where we contrast the average outcome when $p = 1$ versus when $p = 0$.

\medskip

Stochastic interventions are perhaps closest to what can be realistically implemented in many applications. For example, in our criminology example, it's difficult to imagine a world where we could deterministically force probationers to attend or not attend behavioral health services; however, we may well be able to affect probationers' likelihood of attending services (e.g., by providing public transportation stipends, thereby improving the accessibility of services). In this case, we may be interested in measuring the effect of making services more accessible, which can be viewed as a proxy for the effect of increasing $p$ in the Bernoulli trial example. As we will see in Section \ref{sec:incremental}, incremental effects assume a particular stochastic intervention that is similar to the Bernoulli trial example and approximates many interventions that we may expect to see in practice. Thus, before discussing incremental effects specifically, we outline some fundamentals for stochastic interventions.

\medskip

Let $Q(a \mid x)$ denote the probability distribution of treatment for a stochastic intervention, and let $\mathbb{E} ( Y^Q )$ denote the average outcome under this intervention. The quantity $Q(a \mid x)$ is also known as the \textit{intervention distribution}, because it is a distribution specified by an intervention. This intervention can depend on the covariates $x$, and in this case it is a \emph{dynamic} stochastic intervention.  Or, we can have $Q(a \mid x) = Q(a)$ for all $x$, which would be a static (again potentially stochastic) intervention. The quantity $\E ( Y^Q )$ can then be written as a weighted average of the various potential outcomes $Y^a$, with weights given by $Q(a \mid x)$:
\begin{align}\label{stochastic_effect_def}
    \E(Y^Q) = \int_{\mathcal{A}} \int_{\mathcal{X}} \E(Y^a \mid X=x) \ dQ(a \mid x) \ d\Pb(x)
\end{align}
Causal effects under stochastic interventions are often framed as contrasting different $\E(Y^Q)$ for different $Q(a)$ distributions. For example, as stated earlier, the ATE in (\ref{eqn:ate}) can be viewed as contrasting $\E(Y^Q)$ for two different point-mass distributions, where every subject receives treatment or every subject receives control.

\medskip

Under Consistency and Exchangeability (Assumptions \ref{asm:cons} and \ref{asm:exch}, respectively), we can identify (\ref{stochastic_effect_def}) as
\begin{align} \label{stochastic_effect_ident_def}
    \E(Y^Q) = \int_{\mathcal{A}} \int_{\mathcal{X}} \E(Y \mid X = x, A = a) dQ(a \mid x) d\Pb(x)
\end{align} 
where $\mathcal{A}$ denotes the set of all possible treatment values, and $\mathcal{X}$ are all possible covariate values. Note that the notation $dQ(a \mid x)$ acknowledges that the intervention distribution may depend on covariates. In the binary treatment case, this reduces to
\begin{align*}
    \E(Y^Q) = \int_{\mathcal{X}} \Big\{ \E(Y \mid X = x, A = 0) Q(A = 0 \mid x) + \E (Y \mid X = x, A = 1) Q(A = 1 \mid x) \Big\} d\Pb(x).
\end{align*}

Notably, Positivity (Assumption \ref{asm:pos}) may not be  needed to identify (\ref{stochastic_effect_def}), depending on the definition of $Q$. We now turn to incremental propensity score interventions, which is a stochastic intervention that defines an intuitive causal estimand that does not rely on Positivity for identification.

%%%%%%%%%%%%%%%%%%%%%%%%%
%%% IPSIs

\section{Incremental Propensity Score Interventions} \label{sec:incremental}

An \emph{incremental propensity score intervention}, first introduced in \cite{kennedy2019nonparametric}, is a stochastic dynamic intervention in which the odds of receiving treatment under the observed distribution $\Pb$ are multiplied by $\delta$. In other words, this causal effect asks what would happen if everyone's odds of receiving treatment were multiplied by $\delta$.  That is, letting $Q$ denote the intervention distribution:
$$
\delta = \frac{Q(A = 1 \mid X) / (1 - Q(A = 1 \mid X)}{\Pb(A = 1 \mid X) / (1 - \Pb(A = 1 \mid X)} = \frac{\text{odds}_Q (A = 1 \mid X)}{\text{odds}_{\Pb} (A = 1 \mid X)}
$$
This implies that, for $0 < \delta < \infty$, the probability of receiving treatment under $Q$ is
\begin{equation} \label{eq:inc_ps_int}
Q(A = 1 \mid x) \equiv q (x; \delta, \pi) = \frac{\delta \pi (x)}{\delta \pi (x) + 1 - \pi (x)}
\end{equation}
where $\pi(x) = \Pb(A = 1 \mid X = x)$. In our recidivism example, this intervention would shift each probationer's probability of attending a healthcare service. The incremental parameter $\delta$ is user-specified and controls the direction and magnitude of the propensity score shift. It tells us how much the intervention changes the odds of receiving treatment.

\medskip

For example, if $\delta = 1.5$, then the intervention increases the odds of treatment by $50\%$ for everyone. If $\delta = 1$, then we are left with the observation propensity scores, and $q(x; \delta, \pi) = \pi(x)$. As $\delta$ increases from $1$ towards $\infty$, the shifted propensity scores approach $1$, and as $\delta$ decreases from $1$ towards $0$, the shifted propensity scores approach $0$. There are other interventions one might consider, but shifting the odds of treatment is a simple intervention that gives an intuitive explanation for the parameter $\delta$. 

\begin{remark} \label{remark:del_var_x}
It is possible to let $\delta$ depend on $X$, thereby allowing the intervention distribution $Q$ to modify the treatment process differently based on subjects' covariates. This would lead to more nuanced definitions of treatment effects, potentially at the cost of losing straightforward interpretation of the estimated effects. In fact, taking $\delta$ to be constant is not an assumption; it just defines the particular causal estimand that is targeted for inference. 
\end{remark}

\medskip

Incremental propensity score interventions allow us to avoid the tricky issues with Positivity that were discussed in Section \ref{sec:prelim}.  There are two groups of people for whom the Positivity assumption is violated: people who never attend treatment ($\pi = 0$) and people who always attend treatment ($\pi = 1$). Incremental interventions avoid assuming Positivity for these groups because the intervention leaves their propensity score unchanged: It has the useful property that $\pi = 0 \implies q = 0$ and $\pi = 1 \implies q = 1$.

\medskip

We cannot know \textit{a priori} whether Positivity is violated, so this intervention allows us to compute effects on our data without worrying whether Positivity holds. Thus, this intervention differs from the dynamic intervention in Section \ref{sec:prelim}, because we do not make our intervention depend on the propensity score of each individual in our sample. Practically, this means we could still include in our sample people who always attend or never attend treatment; e.g., in our running recidivism example, we could still include individuals who must attend treatment as part of their probation, and the causal effect is still well-defined.  

\begin{remark}
The reader may wonder why we do not consider simpler interventions, such as $q(x; \delta, \pi) = \pi + \delta$ or $q(x; \delta, \pi) = \pi \cdot \delta$.  One reason is that these interventions require the range of $\delta$ to depend on the distribution $\Pb$, because otherwise $q(x; \delta, \pi)$ may fall outside the unit interval. In contrast, for any $\delta$, the incremental propensity score intervention constrains $Q$ so that $0 \leq q(x; \delta, \pi) \leq 1$.
\end{remark}

\begin{remark}
If Positivity holds, incremental interventions nest the ATE as a special case.  If $\pi(x)$ is bounded away from zero and one almost surely, then $\mathbb{E}(Y^Q)$ tends to $\E(Y^1)$ as $\delta \to \infty$ and to $\E(Y^0)$ as $\delta \to 0$. Thus, both $\E(Y^1)$ and $\E(Y^0)$, and thus the ATE, can be approximated by taking $\delta$ to be very large or very small.
\end{remark}

\subsection{Identification} \label{subsec:inc_id}

Under Consistency and Exchangeability (Assumptions  \ref{asm:cons} and \ref{asm:exch}), we can plug the incremental intervention distribution $Q(a \mid x) = q(x; \delta, \pi)^a\{1 - q(x; \delta, \pi)\}^{1 - a}$ into equation \eqref{stochastic_effect_ident_def} to derive an identification expression for $\psi (\delta) = \E \{ Y^{Q (\delta)} \}$, the expected outcome if the treatment distribution is intervened upon and set to $Q(a \mid x)$.

\begin{theorem} \label{thm:id_T1}
\emph{Under Assumptions \ref{asm:cons}-\ref{asm:exch}, and if $\delta \in (0, \infty)$, the incremental effect $\psi(\delta) = \E \{ Y^{Q (\delta)} \}$ for the propensity score intervention as defined in equation (\ref{eq:inc_ps_int}) equals
\begin{align} \label{eq:thm1}
\psi (\delta) = \E \left[ \frac{\delta \pi (X) \mu(X, 1) + \{1 - \pi(X)\} \mu(X, 0)}{\delta \pi(X) + \{ 1 - \pi(X)\}} \right] = \E \left[ \frac{Y(\delta A + 1 - A)}{\delta \pi(X) + \{ 1 - \pi(X)\}} \right]
\end{align}
where $\mu(x, a) = \E(Y \mid X = x, A = a)$.}
\end{theorem}
Theorem \ref{thm:id_T1} offers two ways to link the incremental effect $\psi(\delta)$ to the data generating distribution $\Pb$.  The first involves both the outcome regressions $\mu(x,a)$ and the propensity score $\pi(x)$, and the second just the propensity score. The former is a weighted average of the regression functions $\mu(x, 1)$ and $\mu(x, 0)$, where the weight on $\mu(x, a)$ is given by the fluctuated intervention propensity score $Q(A = 1 \mid x)$ while the latter is a weighted average of the observed outcomes, where the weights are related to the intervention propensity score and depend on the observed treatment. 

\begin{remark}
From the identification expression in equation \eqref{eq:thm1}, one can notice that even if there are subjects for which $\Pb(A = a \mid X) = 0$, so that $\mu(X, a) = \E(Y \mid X, A = a)$ is not defined because of conditioning on a zero-probability event, $\psi(\delta)$ is still well-defined, because those subjects receive zero weight when the expectation over the covariates' distribution is computed. 
\end{remark}

\subsection{Estimation}

Theorem \ref{thm:id_T1} provides two formulas to link the causal effect $\psi(\delta)$ to the data generating distribution $\Pb$. The next step is to estimate $\psi(\delta)$ relying on these identification results.  The first estimator, which we call the ``outcome-based estimator'', includes estimates for both $\mu$ and $\pi$:
\begin{align*}
    \widehat{\psi} (\delta) = \frac{1}{n} \sum_{i=1}^{n} \left[ \frac{\delta \widehat{\pi}(X_i) \widehat{\mu}(X_i, 1) + \{1 - \widehat{\pi}(X_i)\} \widehat{\mu}(X_i, 0)}{\delta \widehat{\pi}(X_i) + \{ 1 - \widehat{\pi}(X_i) \} } \right]
\end{align*}
The second estimator motivated by the identification result is the inverse-probability-weighted (IPW) estimator:
\begin{align*}
    \psihat(\delta) = \frac{1}{n}\sum_{i=1}^n \left[ \frac{Y_i(\delta A_i + 1 - A_i)}{\delta \pihat(X_i) + \{ 1 - \pihat(X_i)\}} \right]
\end{align*}
Both estimators are generally referred to as ``plug-in'' estimators, since they take estimates for $\widehat{\mu}$ or $\widehat{\pi}$ and plug them directly into the identification results.  If we assume we can estimate $\mu$ and $\pi$ with correctly specified parametric models, then both estimators inherit parametric rates of convergence and can be used to construct valid confidence intervals.  

\begin{comment}
When deciding between the outcome-based and inverse-propensity-weighted estimators, one can consider it as trading between bias and variance.  Both estimators require $\widehat{\pi}$ because the intervention shifts the propensity score, but the outcome-based estimator also requires estimating $\widehat{\mu}$.  Thus, the outcome-based estimator may inherit the bias from estimating $\widehat{\mu}$.  On the other hand, the outcome-based estimator is the MLE if both parametric models are correct (TODO: show the MLE proof), so will have lower variance than the IPW estimator.
\end{comment}

\medskip

However, in practice, parametric models are likely to be misspecified.  Thus, we may prefer to estimate the nuisance regression functions $\mu$ and $\pi$ with large, agnostic nonparametric models.  However, if flexible, nonparametric models are used for either plug-in estimators without any correction, the estimators are generally sub-optimal, because they inherit the large bias incurred in estimating the regression functions in large classes. For example, the best possible convergence rate in mean-square-error of an estimator of a regression function that belongs to a H\"{o}lder class of order $\alpha$, essentially a function that is $\alpha$-times differentiable, scales at $n^{-2\alpha / (2\alpha + d)}$, where $n$ is the sample size and $d \geq 1$ is the dimension of the function's input (\cite{tsybakov2009introduction}). This rate is slower than $n^{-1}$ for any $\alpha$ and $d$. Because plug-in estimators with no further corrections inherit this ``slow rate,'' they lose $n^{-1}$ efficiency if the nuisance functions are not estimated parametrically. 

\begin{remark}
We remind readers that the issues regarding plug-in estimators outlined in the previous paragraph are not isolated to incremental interventions; they apply generally to plug-in estimators of functionals.  So, for example, these problems also apply to the outcome-based and IPW estimators for the ATE.
\end{remark}

\medskip

Semiparametric efficiency theory provides a principled way to construct estimators that make a more efficient use of flexibly estimated nuisance functions (\cite{bickel1993efficient, van2000asymptotic, kennedy2016semiparametric, robins2009quadratic, tsiatis2006semiparametric}).\footnote{For a gentle introduction to the use of influence functions in functional estimation, we refer to \cite{fisher2021visually} and the tutorial at \url{http://www.ehkennedy.com/uploads/5/8/4/5/58450265/unc_2019_cirg.pdf}}  Such estimators are based on \textit{influence functions} and are designed for parameters that are ``smooth'' transformations of the distribution $\Pb$. The parameter $\psi(\delta)$ is an example of a smooth parameter. The precise definition of smoothness in this context can be found in Chapter 25 of \cite{van2000asymptotic}. Informally, however, we can note that $\psi(\delta)$ only involves differentiable functions of $\pi$ and $\mu_a$, thereby suggesting that it is a smooth parameter. 

\medskip

If we let $\Phi(\Pb) \in \R$ denote some smooth target parameter, we can view $\Phi(\Pb)$ as a functional acting on the space of distribution functions. One key feature of smooth functionals is that they satisfy a functional analog of a Taylor expansion, sometimes referred to as the \textit{von-Mises expansion}. Given two distributions $\Pb_1$ and $\Pb_2$, the von-Mises expansion of $\Phi(\Pb)$ is
\begin{align*}
\Phi(\Pb_1) - \Phi(\Pb_2) = \int \phi(z; \Pb_1) \{d\Pb_1(z) - d\Pb_2(z)\} + R(\Pb_1, \Pb_2)
\end{align*}
where $\phi(z; \Pb)$ is the (mean-zero) influence function and $R$ is a second-order remainder term. For the purpose of estimating $\Phi(\Pb)$, the above expansion is useful when $\Pb_2 = \Pb$ denotes the true data-generating distribution and $\Pb_1 = \widehat\Pb$ is its empirical estimate. Thus, the key step in constructing estimators based on influence functions is to express the first-order bias of plug-in estimators as an expectation with respect to $\Pb$ of a particular, fixed function of the observations and the nuisance parameters, referred to as the influence function. Because the first-order bias is expressed as an expectation with respect to the data-generating $\Pb$, it can be estimated with error of order $n^{-1}$ simply by replacing $\Pb$ by the empirical distribution. This estimate of the first-order bias can be subtracted from the plug-in estimator, so that the resulting estimator will have second-order bias without any increase in variance. This provides an explicit recipe for constructing ``one-step'' bias-corrected estimators of the form
\begin{align*}
\widehat\Phi(\Pb) = \Phi(\widehat\Pb) + \frac{1}{n} \sum_{i = 1}^n \phi(Z_i; \widehat\Pb)
\end{align*}
If $\phi(Z; \Pb)$ takes the form $\varphi(Z; \Pb) - \Phi(\Pb)$, then the estimator simplifies to $\widehat\Phi(\Pb) = n^{-1} \sum_{i = 1}^n \varphi(Z_i; \widehat\Pb)$. Remarkably, there are other ways of doing asymptotically equivalent bias corrections, for example, through targeted maximum likelihood (\cite{van2006targeted}). 

\medskip

The functional $\psi(\delta)$ satisfies the above von-Mises expansion for 
\begin{align*}
\phi(Z; \Pb) = \frac{\delta \pi(X) \phi_1 (Z) + \{1 - \pi(X)\} \phi_0 (Z)}{\delta \pi(X) + \{1 - \pi(X) \}} + \frac{\delta \gamma(X) \{A - \pi(X)\}}{\{\delta \pi(X) + 1 - \pi(X) \}^2} - \psi(\delta) \equiv \varphi(Z; \Pb) - \psi(\delta)
\end{align*}
where $\gamma(X) = \mu(X, 1) - \mu(X, 0)$ and 
$$
\phi_a (Z) = \frac{\one (A = a)}{\mathbb{P} (A = a \mid X)} \left\{ Y - \mu(X, a) \right\} + \mu(X, a).
$$
To highlight that $\varphi(Z; \Pb)$ depends on $\Pb$ through $\pi$ and $\mu$ and that it also depends on $\delta$, we will write $\varphi(Z; \Pb) \equiv \varphi(Z; \delta, \pi, \mu)$. The influence-function-based one-step estimator of $\psi(\delta)$ is therefore
\begin{align}\label{est:1tmp}
\psihat(\delta) & = \frac1n \sum_{i=1}^{n} \varphi(Z_i; \delta, \pihat, \muhat)
\end{align}

\begin{remark}
The function $\phi_a(Z)$ is the un-centered influence function for the parameter $\E\{\mu(X, a)\}$, and is thus part of the influence function for the ATE under Exchangeability, which can be expressed as $\E\{\mu(X, 1)\} - \E\{\mu(X, 0)\}$. Thus, one may view the first term in the expression for $\psihat(\delta)$ roughly as a weighted mean of the influence functions for the ATE. The second term in the expression arises from the need to estimate $\pi(X)$. 
\end{remark}

It can be advantageous to construct $\pihat$ and $\muhat$ on a sample that is independent from that used to average $\varphi(Z; \delta, \pihat, \muhat)$, such that the observations $Z_i$ in \eqref{est:1tmp} are not used to estimate $\pi$ and $\mu$. The role of the two samples can then be swapped so that two estimates of $\psi(\delta)$ are computed and their average is taken to be the final estimate. This technique, known as cross-fitting, has a long history in statistics (\cite{klaassen1987consistent, bickel1988estimating, robins2008higher, zheng2010asymptotic, chernozhukov2018double}) and is useful to preserve convergence to a Gaussian distribution while avoiding restrictions on the class of  nuisance functions ($\pi$ and $\mu$) and their estimators; see also Remark \ref{remark:cf}.

The full estimator of $\psi(\delta)$ that incorporates sample splitting is detailed in Algorithm \ref{alg:1tp}:
\begin{algorithm} \label{alg:1tp}
Split the data into $K$ folds (e.g., 5), where fold $k \in \{1, \ldots, K\}$ has $n_k$ observations. For each fold $k$:
\begin{enumerate}
    \item Build models $\pihat_{-k}(X)$, $\muhat_{1, -k}(X)$ and $\muhat_{0, -k}(X)$ using the observations \textbf{not} contained in fold $k$.
    \item For each observation $j$ in fold $k$, calculate its un-centered influence function $\varphi(Z_j; \delta, \pihat_{-k} (X_j), \muhat_{-k} (X_j))$ using the models $\pihat_{-k}$ and $\muhat_{-k}$ calculated in the previous step.
    \item Calculate an estimate for $\psi(\delta)$ within fold $k$ by averaging the estimates of the un-centered influence function:
    $$
    \widehat{\psi}_k (\delta) = \frac{1}{n_k}\sum_{j \in k} \varphi\{ Z_j; \delta, \pihat_{-k}(X_j), \muhat_{-k}(X_j) \}
    $$
    \end{enumerate}
    Calculate the final estimate $\psihat(\delta)$ as the average of the $K$ estimates from each fold:
    $$
    \widehat{\psi} = \frac{1}{K}\sum_{k=1}^{K} \widehat{\psi}_k (\delta)
    $$
\end{algorithm}

\medskip
An implementation of the estimator described in Algorithm \ref{alg:1tp} can be found in the the \texttt{R} package \texttt{npcausal} by running the function \texttt{ipsi}. The package can be installed from Github at \url{https://github.com/ehkennedy/npcausal}. 

\subsection{Properties of the estimator}
\begin{comment}
The estimator outlined in Algorithm \ref{alg:1tp} is useful because it allows us to conduct valid inference under mild conditions.  Because it was constructed using the influence function, our estimator does not restrict the nuisance function estimators for $\pihat$ and $\muhat$ to exist in parametric classes.  In fact, the estimators are not even restricted to Donsker classes (a much larger class) because of the sample splitting used in Algorithm \ref{alg:1tp}. 
\end{comment}

\subsubsection{Pointwise inference}\label{sec:ptwise_inf}

Consider the estimator described in Algorithm \ref{alg:1tp}; for each fold $k$, $\pi$ and $\mu$ are estimated using all units except those in fold $k$.  The units outside fold $k$ are independent of the units in fold $k$, and the units in fold $k$ are used to calculate the sample average. This means that, under mild regularity conditions, the estimators from each fold have variances of order $n^{-1}$ conditional on the training sample, as they are sample averages of independent observations. Conditioning on the training sample, the bias can be upper bounded by a multiple of
\begin{align}\label{eq:bias}
B \equiv \int \{\pi(x) - \pihat(x)\}^2 d\Pb(x) + \int \{\pi(x) - \pihat(x)\}\max_a\{\mu_a(x) - \muhat_a(x)\} d\Pb(x)
\end{align}
This bias term is second-order. Thus, remarkably, the bias can be $o_\Pb(n^{-1/2})$, and thus of smaller order than the standard error, even if $\pi$ and $\mu$ are estimated at slower rates. If the bias is asymptotically negligible ($o_{\Pb} (n^{-1/2})$), then 
$$
\frac{\sqrt{n}\{\psihat(\delta) - \psi(\delta)\}}{\sigma(\delta)} = \frac{\sqrt{n}\{\widetilde\psi(\delta) - \psi(\delta)\}}{\sigma(\delta)} + o_\Pb(1) \indist N(0, 1),
$$ 
where $\sigma^2(\delta) = \var\{\varphi(Z; \delta, \pi, \mu)\}$ and $\widetilde{\psi}(\delta) = n^{-1} \sum_i \varphi(Z_i; \delta, \pi, \mu)$. Given a consistent estimator of the variance $\widehat\sigma^2(\delta)$, by Slutsky's theorem, it holds that
% The variance can be consistently estimated by
% \begin{align*}
% \widehat\sigma^2(\delta) = \frac{1}{n}\sum_{i=1}^n\left\{ \varphi(Z_i; \delta, \pihat, \muhat) - \psihat(\delta) \right\}^2
% \end{align*}
\begin{align} \label{convergence_ptwise_1tp}
\frac{\sqrt{n}\{\psihat(\delta) - \psi(\delta)\}}{\widehat\sigma(\delta)} \indist N(0, 1).
\end{align}
Revisiting the bias term, we can see that it can be $o_\Pb(n^{-1/2})$ even when the nuisance regression functions are estimated nonparametrically with flexible machine learning methods. By Cauchy-Schwarz,
\begin{align*}
|B| \leq \Norm{\pi - \pihat}^2 + \Norm{\pi - \pihat}\max_a\Norm{\mu_a - \muhat_a},
\end{align*}
where $\Norm{f}^2 = \int f^2(z) d\Pb(z)$. Therefore, it is sufficient that the product of the integrated MSEs for estimating $\pi$ and $\mu_a$ converge to zero faster than $n^{-1/2}$ rates to ensure that the bias is asymptotically negligible. This can happen, for instance, if both $\pi$ and $\mu_a$ are estimated faster than $n^{-1/4}$ rates, which is possible in nonparametric models under structural conditions such as sparsity or smoothness. 

\medskip

The convergence statement in \eqref{convergence_ptwise_1tp} provides an asymptotic Wald-type $(1-\alpha)$-confidence interval for $\psi(\delta)$,
\begin{equation} \label{eq:pt_ci}
\psihat(\delta) \pm z_{1-\alpha/2} \cdot \frac{\widehat\sigma(\delta)}{\sqrt{n}},
\end{equation}
where $z_\tau$ is the $\tau$-quantile of a standard normal.  This confidence interval enables us to conduct valid inference for $\psi (\delta)$ for a particular value of $\delta$. 

\begin{remark} \label{rem:smooth}
If $\pi$ and $\mu_a$ are in a H\"{o}lder class of order $\alpha$, then the best estimators in the class have MSEs of order $n^{-2\alpha / (2\alpha + d)}$, where $d$ is the dimension of $X$ (\cite{tsybakov2009introduction}). Therefore, the bias term would be of order $o_\Pb(n^{-1/2})$ whenever $d < 2\alpha$. Similarly, if $\mu_a$ follows a $s$-sparse linear model and $\pi$ a $s$-sparse logistic model, \cite{farrell2015robust} shows that $s^2 (\log d)^{3 + 2 \delta} = o(n)$, for some $\delta > 0$, is sufficient for the the bias to be asymptotically negligible.   
\end{remark}

\begin{remark}\label{remark:cf}
To see why estimating the nuisance functions on an independent sample may help, consider the following expansion for $\psihat(\delta)$ defined in \eqref{est:1tmp}:
\begin{align*}
\psihat(\delta) - \psi(\delta) & = (\Pn - \Pb)\{\varphi(Z; \delta, \pihat, \muhat) - \varphi(Z; \delta, \pi, \mu)\} + (\Pn - \Pb)\{\varphi(Z; \delta, \pi, \mu)\} \\
& \quad + \Pb\{\varphi(Z; \delta, \pihat, \muhat) - \varphi(Z; \delta, \pi, \mu)\}
\end{align*}
where we used the shorthand notation $\Pn\{g(Z)\} = n^{-1}\sum_{i = 1}^n g(Z_i)$ and $\Pb\{g(Z)\} = \int g(z) d\Pb(z)$. The second term will be of order $O_\Pb(n^{-1/2})$ by the Central Limit Theorem, whereas the third term can be shown to be second order (in fact, upper bounded by a multiple of $B$ in \eqref{eq:bias}) by virtue of $\varphi(Z; \delta, \pi, \mu)$ being the first-order influence function. 

\medskip 

Cross-fitting helps with controlling the first term. If $g(Z_i) \ind g(Z_j)$, then $(\Pn - \Pb)\{g(Z)\} = O_\Pb(\Norm{g(Z)} / \sqrt{n})$, where $\Norm{g(Z)}^2 = \int g^2(z) d\Pb(z)$. If $\pihat$ and $\muhat$ are computed on a separate training sample, then, given that separate sample, the first term is 
\begin{align*}
(\Pn - \Pb)\{\varphi(Z; \delta, \pihat, \muhat) - \varphi(Z; \delta, \pi, \mu)\} = O_\Pb \left\{ \frac{ \Norm{\varphi(Z; \delta, \pihat, \muhat) - \varphi(Z; \delta, \pi, \mu)}}{\sqrt{n}} \right\}
\end{align*}
Therefore, because of cross-fitting, the first-term is $o_\Pb(n^{-1/2})$ as long as $\Norm{\varphi(Z; \delta, \pihat, \muhat) - \varphi(Z; \delta, \pi, \mu)} = o_\Pb(1)$.  This is a very mild consistency requirement. In the absence of cross-fitting, the first term would not be a centered average of independent observations. It would still be $o_\Pb(n^{-1/2})$ if $\Norm{\varphi(Z; \delta, \pihat, \muhat) - \varphi(Z; \delta, \pi, \mu)} = o_\Pb(1)$ and the process $\{\sqrt{n}(\Pn - \Pb) \{\varphi(Z; \delta, \overline\pi, \overline\mu)\}: \overline\pi \in \mathcal{P}, \overline\mu \in \mathcal{M}\}$ is stochastically equicontinuous, where $\mathcal{P}$ and $\mathcal{M}$ are the classes where $\pihat$ and $\muhat$, together with their limits, live, respectively. This would require $\mathcal{P}$ and $\mathcal{M}$ to satisfy specific ``Donsker-type'' conditions.  For example, $\mathcal{P}$ and $\mathcal{M}$ could not be overly complex in the sense of having infinite uniform entropy integrals. Therefore, cross-fitting allows us to avoid these Donsker-type conditions on $\mathcal{P}$ and $\mathcal{M}$ and still have the first term in the decomposition above be $o_{\Pb}(n^{-1/2})$.  For more discussion, please see Lemma 1 in \cite{kennedy2020sharp} and Chapter 2 in \cite{van1996weak}. 
\end{remark}

\subsubsection{Uniform inference} \label{sec:unif_inf}

The Wald-type confidence interval in equation \eqref{eq:pt_ci} is valid for a particular value of $\delta$.  For example, if we only cared about what might happen if we multiply the odds of attending treatment by two, then we can use the estimator in Algorithm \ref{alg:1tp} with $\delta = 2$ and the confidence interval in \eqref{eq:pt_ci}.  However, in almost any applied analysis, we care about comparing different levels of shifts $\delta$.  In the most basic example, we would care at least about $\widehat{\psi} (\delta = 1)$ and $\widehat{\psi}(\delta \neq 1)$ so we can compare the observational data to an intervention.  In general, it is useful to vary $\delta \in \mathcal{D} \subseteq (0, \infty)$ and trace a curve $\psihat(\delta)$ as a function of $\delta$.  In this section we outline the convergence of $\psihat(\delta)$ to $\psi(\delta)$ in the function space $\ell^\infty(\mathcal{D})$, i.e., the space of all bounded functions on $\mathcal{D}$.

\medskip

If $\mathcal{D} \subseteq (0, \infty)$, then the function $\delta \mapsto \varphi(Z; \delta, \pi, \mu)$ belongs to the class of Lipschitz functions, which is sufficiently small to satisfy Donsker conditions. As such, the quantity $\sqrt{n}\{\widetilde{\psi}(\delta) - \psi(\delta)\} / \sigma(\delta)$ converges to a mean-zero Gaussian process in $\ell^\infty(\mathcal{D})$:
\begin{align*}
\frac{\sqrt{n}\{\widetilde{\psi}(\delta) - \psi(\delta)\}}{ \sigma(\delta)}\indist G(\delta)
\end{align*}
where $G(\cdot)$ is a mean-zero Gaussian process with covariance
\begin{align*}
\E\{G(\delta_1)G(\delta_2)\} = \E\left[\left\{ \frac{\varphi(Z; \delta_1, \pi, \mu) - \psi(\delta_1)}{\sigma(\delta_1)}\right\}\left\{ \frac{\varphi(Z; \delta_2, \pi, \mu) - \psi(\delta_2)}{\sigma(\delta_2)}\right\}\right]
\end{align*}
This means that, under essentially the same conditions that guarantee asymptotic normality at a fixed value of $\delta$, it is also the case that any finite linear combination $\psi(\delta_1), \psi(\delta_2), \ldots, \psi(\delta_m)$ is asymptotically distributed as a multivariate Gaussian. 

\medskip

Establishing sufficient conditions to achieve convergence to a Gaussian process allows us to conduct uniform inference across many $\delta$'s; i.e., we can conduct inference for $\psi(\delta)$ for many $\delta$'s without issues of multiple testing. In particular, we can construct confidence bands around the curve $\psihat(\delta)$ that covers the true curve with a desired probability across all $\delta$.  The bands can be constructed as $\psihat(\delta) \pm \widehat{c}_{\alpha} \widehat\sigma(\delta)$, where $\widehat{c}_{\alpha}$ is an estimate of the $(1 - \alpha)$-quantile of $\sup_{\delta \in \mathcal{D}} \sqrt{n}|\psihat(\delta) - \psi(\delta)| / \widehat\sigma(\delta)$.  We can estimate the supremum of Gaussian processes using the multiplier bootstrap, which is computationally efficient. We refer the reader to \cite{kennedy2019nonparametric} for full details. 

\medskip

With the uniform confidence interval, we can also conduct a test of no treatment effect.  If the treatment has no effect on the outcome, then $Y \ind A \mid X$ and $\psi(\delta) = \E(Y)$ under Exchangeability. Given the uniform confidence band, the null hypothesis of no incremental intervention effect
$$
H_0 : \psi(\delta) = \E(Y) \text{ for all } \delta \in \mathcal{D},
$$
can be tested by checking whether a $(1 - \alpha)$ band contains a straight horizontal line over $\mathcal{D}$. That is, we reject $H_0$ at level $\alpha$ if we {cannot} run a straight horizontal line through the confidence band. We can also compute a p-value as
$$
\widehat{p} = \sup \left[ \alpha: \inf_{\delta \in \mathcal{D}} \left\{ \widehat{\psi}(\delta) + \widehat{c}_{\alpha} \widehat{\sigma}(\delta) / \sqrt{n} \right\} \geq \sup_{\delta \in \mathcal{D}} \{ \widehat{\psi} (\delta) - \widehat{c}_{\alpha} \widehat{\sigma} (\delta) / \sqrt{n} \} \right].
$$
Geometrically, the p-value corresponds to the $\alpha$ at which we can no longer run a straight horizontal line through our confidence band.  At $\alpha = 0$ we have an infinitely wide confidence band and we always fail to the reject $H_0$.  Increasing $\alpha$ corresponds to a tightening confidence band. In Section \ref{sec:example} we give a visual example of how to conduct this test in practice. But first, we discuss incremental propensity score interventions when treatment is time-varying.

%%%%%%%%%%%%%%%%%%%%%%%%%
%%% Time-varying

\section{Time-varying treatments} \label{sec:time}

\subsection{Notation}

In this section, we extend the prior results from the one time point case to the time-varying setup. In the time-varying setup we consider $n$ samples $(Z_1, ..., Z_n)$ IID from some distribution $\Pb$ over $T$ time-points such that
$$
Z = (X_1, A_1, X_2, A_2, ..., X_T, A_T, Y)
$$
where $X_t$ are time-varying covariates, $A_t$ is time-varying treatment, and $Y$ is the outcome. In many studies, only a few covariates are time-varying and most are measured at baseline; all baseline covariates are included in $X_1$.  Additionally, in some studies there are time-varying outcome variables $Y_t$; in that case, we can include the time-varying outcomes as part of the time-varying covariates $X_{t+1}$. We consider potential outcomes $Y^{\overline{a}_T}$, which is the outcome we would have observed if the treatment regime $\overline{A}_T = \overline{a}_T$, where $\overline{A}_t$ is the history of $A_t$ from period 1 until $t$ and $H_t = (\overline{A}_{t-1}, \overline{X}_t)$ is the full history up until treatment in period $t$.

\medskip

First, we make two time-varying assumptions that are analogous to Assumptions \ref{asm:cons}-\ref{asm:exch}:

\begin{assumption} \label{asm:cons_time} (Consistency). $Y = Y^{\overline{a}_T}$ if $\overline{A}_T = \overline{a}_T$. \end{assumption}
\begin{assumption} \label{asm:exch_time} (Sequential Exchangeability). $A_t \ind Y^{\overline{a}_T} \mid H_t$. \end{assumption}
Usually, Positivity, is also assumed.  As we saw in Section \ref{sec:incremental}, Positivity will not be required for incremental interventions, but we will state it here:
\begin{assumption} \label{asm:pos_time} (Positivity). There exists $\epsilon > 0$ such that $\Pb( \epsilon < \Pb(\overline{A}_T = \overline{a}_T \mid X) < 1 - \epsilon) = 1$ for any treatment regime $\overline{a}_T$. \end{assumption}
Consistency says that the observed outcome under a treatment sequence equals the potential outcome under that treatment sequence.  Sequential Exchangeability  says the outcome \emph{at each time point} is effectively randomized within covariate history strata.  Positivity says that everyone has a non-zero chance of taking each treatment regime.  We discussed each assumption in detail in Section \ref{subsec:inc_id} and showed how these assumptions might be violated.

\subsection{Marginal structural models}

In time-varying causal inference, or generally when the treatment can take many values, a tension arises between the number of possible interventions and how well we can estimate them.  For example, a binary treatment administered at $T$ time points leads to $2^T$ possible treatment regimes and as many possible interventions to estimate. Without further restrictions, the number of parameters and the sample size required to estimate them grow exponentially with $T$. 

\medskip

A popular approach to reduce the number of parameters is to assume that the expected potential outcomes under different interventions vary smoothly. For instance, one can specify a model of the form 
\begin{align*}
\E (Y^{\overline{a}_T}) = m(\overline{a}_T; \beta),
\end{align*}
where $m(\overline{a}_T; \beta)$ is a function known up to a finite-dimensional parameter $\beta \in \R^p$. This model, termed a \emph{marginal structural model} (MSM), imposes a parametric structure on the relationship between the potential outcomes and the treatment regime.  A simple example allows the potential outcomes only to depend on the cumulative treatment attendance:
$$
m(\overline{a}_T; \beta) = \beta_0 + \beta_1 \left( \sum_{t=1}^{T} a_t \right).
$$
A popular approach to estimate $\beta$ is through the moment condition
\begin{equation} \label{eq:ipw_msm}
\E \left[ h(\overline{A}_T) W(Z) \left\{Y - m(\overline{A}_T; \beta) \right\} \right] = 0
\end{equation}
where 
$$
W(Z) = \frac{1}{\prod_{t=1}^{T} d\Pb (A_t \mid H_t)}
$$
and $h$ is some arbitrary function of the treatment.  This moment condition suggests an inverse-propensity weighted estimator, where we estimate $\widehat{\beta}$ by solving the empirical analog of \eqref{eq:ipw_msm}:
$$
\frac{1}{n}\sum_{i = 1}^n \left\{ h(\overline{A}_{T, i}) \widehat{W}(Z_i) \{Y_i - m(\overline{A}_{T, i}; \widehat{\beta})\} \right\} = 0.
$$
We can also estimate $\beta$ using a doubly robust version of the moment condition (\cite{bang2005doubly}).

\medskip

MSMs are a major advance towards performing sound causal inference in time-varying settings. However, there are two important issues they cannot easily resolve. First, while specifying a model for $\E(Y^{\overline{a}_T})$ is a less stringent requirement than, say, specifying a model for the outcome given treatment and covariates, it can still lead to biased estimates if $m(\overline{a}_T; \beta)$ is not correctly specified. Second, identifying and estimating $\beta$ still relies on Positivity, which is unlikely to be satisfied when there are many time points or treatment values. In fact, even if Positivity holds by design as it would in an experiment, we are unlikely to observe all treatment regimes in a given experiment simply because the number of possible treatment regimes grows exponentially with the number of time points.  This poses a computational challenge even when Positivity holds, because the product of densities in the denominator of (\ref{eq:ipw_msm}) may be very small, resulting in an unstable estimate of $\beta$.  

\medskip

There are ways to mitigate the two drawbacks outlined above, but neither issue can be completely fixed. First, one does not necessarily need to assume the MSM is correctly specified. Instead, one can use the ``working model'' approach and estimate a projection of $\E(Y^{\overline{a}_T})$ onto the MSM $m(\overline{a}_T; \beta)$.  In this approach, one estimates $\beta$ as the parameter that yields the best approximation of the causal effect in the function class described by $m(\overline{a}_T; \beta)$ (\cite{neugebauer2007nonparametric}). This approach is beneficial because it allows for model-free inference: we can construct valid confidence intervals for the projection parameter $\beta$ regardless of whether the MSM is correctly specified.  However, we are still only estimating a projection; so, this approach can be of limited practical relevance if the model is grossly misspecified and the projection onto the model has very little bearing on reality. 

\medskip

Meanwhile, to mitigate in-sample Positivity violations, a popular approach is to ``stabilize'' the propensity scores by replacing $W(Z)$ in \eqref{eq:ipw_msm} with 
\begin{align*}
\widetilde{W}(Z) = \prod_{t=1}^T \frac{\Pb(A_t \mid \overline{A}_{t-1})}{\Pb(A_t \mid H_t)}.
\end{align*}
See \cite{cole2008constructing} and \cite{talbot2015cautionary} for discussions of the advantages and disadvantages of using the stabilized weights $\widetilde{W}(Z)$ instead of $W(Z)$. Another approach is propensity score trimming, which we also discussed in the one time point case.  However, as we discussed in Section \ref{sec:prelim}, these ad-hoc approaches are not guaranteed to solve Positivity violations. 

\begin{remark}
MSMs struggle with Positivity violations in the same way traditional ATE estimators struggle with Positivity violations in the one time point case.  In Section \ref{sec:prelim}, we outlined how these Positivity issues lead naturally to dynamic and stochastic interventions.  A similar logic applies here, although perhaps more urgently, because Positivity violations occur in time-varying analyses more frequently.
\end{remark}

\begin{remark}
MSMs have also been criticized because, even if the model is correctly specified and Positivity is not violated, the causal effect of the treatment may be hard to interpret and visualize since it may be implicitly encoded in a complicated model $m(\overline{a}_T; \beta)$.   While this is a solvable problem, we mention it so the reader might compare the difficulty of interpreting a complicated model $m(\overline{a}_T; \beta)$ with the ease of understanding the delta curve graphs in Section \ref{sec:example}.
\end{remark}

\subsection{Time-varying Incremental Effects}

A \emph{time-varying incremental propensity score intervention} takes the same form as the one time-point intervention. So, at time $t$, we multiply the odds of receiving treatment by $\delta$.  Thus, we have a stochastic dynamic intervention $Q_t (A_t \mid H_t)$ where:
\begin{equation} \label{eq:int_time}
    Q_t(A_t = 1 \mid H_t) \equiv q_t (H_t; \delta, \pi_t) = \frac{\delta \pi_t (H_t)}{\delta \pi_t (H_t) + 1 - \pi_t (H_t)} \text{  for } 0 < \delta < \infty.
\end{equation}
for $\pi_t(H_t) = \Pb(A_t = 1 \mid H_t)$. This is the same intervention as in equation \eqref{eq:inc_ps_int}, just with time subscripts added and conditioning on all the past up to treatment in time $t$ (i.e., $H_t$).  There are two main differences between the intervention \eqref{eq:int_time} and the incremental intervention described in Section \ref{sec:incremental}:
\begin{enumerate}
    \item This intervention happens over every time period from $t = 1$ to $T$, the end of the study.
    \item This intervention requires the time-varying analogs of Consistency and Exchangeability, Assumptions \ref{asm:cons_time} and \ref{asm:exch_time}.
\end{enumerate}
Despite these differences, most of the machinery developed in Section \ref{sec:incremental} applies here. The intuition about what happens when we shift $\delta \to 0$ or $\delta \to \infty$ is the same.  Unfortunately, the results and proofs look much more imposing at first glance, but that is due to the time-varying nature of the data, not any change in the incremental intervention approach.

\medskip

The incremental approach is quite different from the MSM approach.  The incremental intervention is a stochastic dynamic intervention that shifts propensity scores in each time period, whereas MSMs describe a static intervention for what would happen if everyone took treatment $\overline{a}_T$. Consequently, the incremental intervention framework does not require us to assume Positivity (Assumption \ref{asm:pos_time}) or a parametric model  $m(\overline{A}_T; \beta)$, whereas MSMs require both.

\begin{remark} \label{remark:var_del_t}
The time-varying incremental intervention can actually allow $\delta$ to vary over $t$, but we omit this generalization for ease of exposition and interpretability. In other words, in equation \eqref{eq:int_time} we could use $\delta_t$ instead of $\delta$ and allow $\delta_t$ to vary with $t$.  Whether allowing $\delta$ to vary with time is useful largely depends on the context.  In some applications it may be enough to study interventions that keep $\delta$ constant across time. On the other hand, one can imagine interventions whose impact varies over time; for example, some encouragement mechanism might have an effect that decays toward zero with time. Either way, the theory and methodology presented here would remain valid.
\end{remark}

\subsection{Identification}

The following theorem is a generalization of Theorem \ref{thm:id_T1}; it shows that the mean counterfactual outcome under the incremental intervention $\left\{ q_t (H_t; \delta, \pi_t) \right\}_{t=1}^{T}$ is identified under Assumptions \ref{asm:cons_time}-\ref{asm:exch_time}.

\begin{theorem} \label{thm:id_inc_time}
\emph{(Theorem 1, \cite{kennedy2019nonparametric}) \\
Under Assumptions \ref{asm:cons_time}-\ref{asm:exch_time}, for $\delta \in (0, \infty)$, the incremental propensity score effect $\psi \left( \delta \right)$ equals
\begin{equation} \label{eq:plugin_id}
    \psi \left(\delta \right) = \sum_{\overline{a}_T \in \mathcal{A}^T} \int_{\mathcal{X}} \mu(h_T, a_T) \prod_{t=1}^{T} \frac{a_t \delta \pi_t (h_t) + (1 - a_t) \{ 1 - \pi_t (h_t) \}}{\delta \pi_t (h_t) + 1 - \pi_t (h_t)} d\Pb(x_t \mid h_{t - 1}, a_{t - 1})
\end{equation}
and 
\begin{equation}
    \psi \left( \delta \right) = \E \left\{ Y \prod_{t=1}^{T} \frac{\delta A_t + 1 - A_t}{\delta \pi_t (H_t) + 1 - \pi_t (H_t)} \right\}
\end{equation}
where $\mathcal{X} = \mathcal{X}_1 \times \cdots \times \mathcal{X}_T$ and $\mu (h_T, a_T) = \E[Y \mid H_T = h_T, A_T = a_t]$.
}
\end{theorem}
To see why the identification formula in equation \eqref{eq:plugin_id} arises, notice that with generic interventions $dQ(a_t \mid h_t)$ we can extend equation \eqref{stochastic_effect_ident_def} to multiple time points with
$$
\E(Y^Q) = \int_{\mathcal{A}_1 \times \cdots \times \mathcal{A}_T} \int_{\mathcal{X}_1 \times \cdots \times \mathcal{X}_T} \mu(h_T, a_T) \prod_{t=1}^{T} dQ(a_t \mid h_t) d\Pb(x_t \mid h_{t - 1}, a_{t - 1}),
$$
which follows by Robins' g-formula (\cite{robins1986new}) and replacing the general treatment process with a generic stochastic intervention $dQ(a_t \mid h_t)$.  Then, we can replace the generic $dQ(a_t \mid h_t)$ with $a_t q_{t}(H_t; \delta, \pi_t) +(1 - a_t)\{1 - q_{t}(H_t; \delta, \pi_t)\}$ and $\int_{\mathcal{A}_1 \times \cdots \times \mathcal{A}_T}$ with $\sum_{\overline{a}_T \in \mathcal{A}_T}$. The full proof is shown in the appendix of \cite{kennedy2019nonparametric}.  Just like in the $T = 1$ case, $\psi(\delta)$ is well-defined even if $\pi_t(h_t) = 0$ or 1 for some $h_t$. 

\subsection{Estimation}

In Section \ref{sec:incremental}, we presented two ``plug-in'' estimators, briefly reviewed the theory of influence functions, and presented an influence-function-based estimator that allows for nonparametric nuisance function estimation. In the multiple time point case, the parameter $\psi(\delta)$ is still smooth enough to possess an influence function. Thus, the discussion in Section \ref{sec:incremental} also applies here, although the notation becomes more involved.  

\medskip

Again, there are two plug-in estimators.  The outcome-based g-computation style estimator is motivated by equation \eqref{eq:plugin_id} in Theorem \ref{thm:id_inc_time}, and can be constructed with
$$
\psi \left(\delta \right) = \sum_{\overline{a}_T \in \mathcal{A}^T} \int_{\mathcal{X}} \widehat{\mu}(h_T, a_T) \prod_{t=1}^{T} \frac{a_t \delta \widehat{\pi}_t (h_t) + (1 - a_t) \{ 1 - \widehat{\pi}_t (h_t) \}}{\delta \widehat{\pi}_t (h_t) + 1 - \widehat{\pi}_t (h_t)} d\widehat{\Pb}(x_t \mid h_{t - 1}, a_{t - 1})
$$
where $\mathcal{X} = \mathcal{X}_1 \times \cdots \times \mathcal{X}_T$.  Similarly, the IPW estimator can be constructed with:
$$
\widehat{\psi} (\delta) = \frac{1}{n} \sum_{i=1}^{n} \left\{ Y_i \prod_{t=1}^{T} \frac{(\delta A_{t,i} + 1 - A_{t,i})}{\delta \pihat_t (H_{t,i}) + 1 - \pihat_t (H_{t,i})} \right\}.
$$
As before, both estimators inherit the convergence rates of the nuisance function estimators for $\pihat_t$ and $\muhat$, and only attain parametric rates of convergence with correctly specified parametric models for \textbf{every} nuisance function. As with the single time-point case, we may be skeptical as to whether specifying correct parametric models is possible, and this motivates an influence-function-based approach. Theorem 2 in \cite{kennedy2019nonparametric} derives the influence function for $\psi(\delta)$. 

\begin{theorem} \label{thm:eif_time}
\emph{(Theorem 2, \cite{kennedy2019nonparametric})} The un-centered efficient influence function for $\psi \left( \delta \right)$ in a nonparametric model with unknown propensity scores is given by
\begin{align*}
    \varphi(Z; \eta, \delta) &\equiv \sum_{t=1}^T \left[ \frac{A_t \{ 1 - \pi_t (H_t) \} - (1 - A_t) \delta \pi (H_t)}{\delta / (1 - \delta) } \right] \times \\
    &\hspace{0.45in} \left[ \frac{\delta \pi_t (H_t) m_t (H_t, 1) + \{1 - \pi_t (H_t) \} m_t (H_t, 0) }{\delta \pi_t (H_t) + 1 - \pi_t (H_t)} \right] \times \left\{ \prod_{s=1}^{t} \frac{\delta A_s + 1 - A_s}{\delta \pi_s (H_s) + 1 - \pi_s (H_s)} \right\} \\
    &+ \prod_{t=1}^{T} \frac{(\delta A_t + 1 - A_t) Y}{\delta \pi_t (H_t) + 1 - \pi_t (H_t)} 
\end{align*}
where $\eta = \{m_1, \ldots, m_T, \pi_1, \ldots, \pi_T\}$, $m_T (h_T, a_T) = \mu(h_T, a_T)$ and for $t = 0, ..., T - 1$:
$$
m_t (h_t, a_t) = \int_{\mathcal{R}_t} \mu(h_T, a_T) \prod_{s=t+1}^{T} [a_sq_s (H_s; \delta, \pi_s) + (1 - a_s)\{1 - q_s (H_s; \delta, \pi_s)\}] d\Pb (x_s \mid h_{s-1}, a_{s-1})
$$
with $\mathcal{R}_t = (\mathcal{H}_T \times \mathcal{A}_T) \setminus \mathcal{H}_t$. 
\end{theorem}
\begin{comment}
\begin{remark}
$\varphi(Z; \eta, \delta)$ is the un-centered efficient influence function for $\psi(\delta)$ because $\E\{\varphi(Z; \eta, \delta)\} = \psi(\delta)$. This is shown in \cite{kennedy2019nonparametric}.  $\varphi(Z; \eta, \delta)$ is \textbf{efficient} because it has minimum variance out of all possible influence functions.
\end{remark}
\end{comment}
The following influence-function-based estimator is a natural consequence of Theorem \ref{thm:eif_time}:
\begin{align*}
    \widehat\psi(\delta) = \frac{1}{n}\sum_{i = 1}^n \varphi(Z_i; \widehat\eta, \delta)
\end{align*}
Similarly to the $T = 1$ case, this estimator is optimal with nonparametric models under certain smoothness or sparsity constraints.   Again, it is advantageous to construct this estimator using cross-fitting, since it allows fast convergence rates of $\widehat\psi(\delta)$ without imposing Donsker-type conditions on the estimators of $\eta$.  The detailed algorithm is provided as Algorithm 1 of \cite{kennedy2019nonparametric}. For intuition, the reader can imagine the algorithm as an extension of Algorithm \ref{alg:1tp} in this chapter, where in step 1 we estimate all the nuisance functions in Theorem \ref{thm:eif_time}.  However, unlike Algorithm \ref{alg:1tp}, we can estimate $m_t$ recursively backwards though time, and this is outlined in Remark 2 and Algorithm 1 from \cite{kennedy2019nonparametric}. The \texttt{ipsi} function in the \texttt{R} package \texttt{npcausal} can be used to calculate incremental effects in time-varying settings. 

\subsection{Inference}

As in the $T = 1$ case, we provide both pointwise and uniform inference results. The theory is essentially the same as in the one time-point case, but the conditions to achieve convergence to a Gaussian distribution or a Gaussian process need to be adjusted to handle multiple time points. 

\medskip

Theorem 3 in \cite{kennedy2019nonparametric} states that, under mild regularity conditions, if 
\begin{align*}
\left( \sup_{\delta \in \mathcal{D}} \Norm{\widehat{m}_{t, \delta} - m_{t, \delta}} + \Norm{\widehat{\pi}_t - \pi_t} \right) \Norm{\widehat{\pi}_s - \pi_s} = o_{\Pb} (n^{-1/2}) \quad
\text{for } s \leq t \leq T
\end{align*}
then
\begin{align*}
\frac{\sqrt{n}\{\widehat{\psi}(\delta) - \psi(\delta)\}}{\widehat{\sigma}(\delta)} \indist \mathbb{G}(\delta)
\end{align*}
in $l^{\infty}(\mathcal{D})$, where $\widehat\sigma(\delta)$ is a consistent estimate of $\var\{\varphi(Z; \eta, \delta)\}$ and $\mathbb{G}(\cdot)$ is a mean-zero Gaussian process with covariance $\E\{ \mathbb{G}(\delta_1) \mathbb{G}(\delta_2) \} = \E \{ \widetilde{\varphi} (Z; \eta, \delta_1) \widetilde{\varphi}(Z; \eta, \delta_2) \}$ and $\widetilde{\varphi}(z; \eta, \delta) = \{\varphi(z; \eta, \delta) - \psi(\delta)\} / \sigma(\delta)$. In particular, for a given fixed value of $\delta$, we have
\begin{align*}
\frac{\sqrt{n}\{\widehat{\psi}(\delta) - \psi(\delta)\}}{\widehat{\sigma}(\delta)} \indist N(0, 1).
\end{align*}
The main assumption underlying this result is that the product of the $L_2$ errors for estimating $m_t$ and $\pi_t$ is of smaller order than $n^{-1/2}$. This is essentially the same small-bias condition discussed in Section \ref{sec:ptwise_inf} for the $T = 1$ case, adjusted to handle multiple time points. This requirement is rather mild, because $m_t$ and $\pi_t$ can be estimated at slower-than-$\sqrt{n}$ rates, say $n^{-1/4}$ rates, without affecting the efficiency of the estimator. As discussed in Remark \ref{rem:smooth}, $n^{-1/4}$ rates are attainable in nonparametric models under smoothness or sparsity constraints. 

\medskip

Finally, the convergence statements above allow for straightforward pointwise and uniform inference as discussed in Sections \ref{sec:ptwise_inf} and \ref{sec:unif_inf}. In particular, we can construct a Wald-type confidence interval 
$$
\psihat(\delta) \pm z_{1 - \alpha / 2} \frac{\widehat\sigma(\delta)}{\sqrt{n}}
$$
at each $\delta$ using the algorithm outlined above.  We can also create uniformly valid confidence bands covering $\delta \mapsto \psi(\delta)$ via the multiplier bootstrap and use the bands to test for no treatment effect, as in Section \ref{sec:unif_inf}. We refer to Sections 4.3 and 4.4 in \cite{kennedy2019nonparametric} for additional technical details.

%%%%%%%%%%%%%%%%%%%%%%%%%%%%
%%% Example

\section{Example Analysis} \label{sec:example}

\begin{figure}[H]
    \centering
    \includegraphics[width=5in]{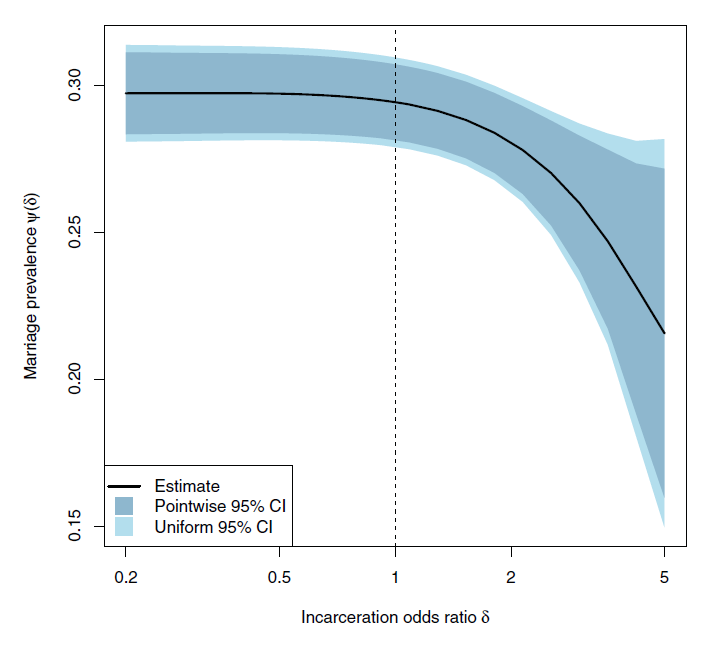}
    \caption{Estimated marriage prevalence $10$ years post-baseline, if the incarceration odds were multiplied by factor $\delta$, with pointwise and uniform $95$\% confidence bands}
    \label{fig:delta_curve}
\end{figure}

In this section, we show an example curve that one could obtain to describe incremental effects.  One could obtain Figure \ref{fig:delta_curve} (Figure 3 from \cite{kennedy2019nonparametric}) by running the \texttt{ipsi} function in the \texttt{npcausal} package.  \cite{kennedy2019nonparametric} reanalyzed a dataset on the effects of incarceration on marriage prevalence. For this figure, they used 10 years of data  (2001-2010) for 4781 individuals from the National Longitudinal Survey of Youth.  The dataset contains baseline covariates like demographic information and delinquency indicators, as well as many time-varying variables such as employment and earnings.  They defined the outcome $Y$ as whether someone was married at the end of the study (i.e., in 2010), and defined the treatment $A_t$ as whether someone was incarcerated in a given year.

\medskip

Figure \ref{fig:delta_curve} shows their estimated $\psi(\delta)$ curve.  The x-axis is the odds multiplier $\delta$, which ranges from $0.2$ to $5$.  Because $\delta$ is a multiplier, this range is a natural choice ($0.2 = \frac{1}{5}$).  The y-axis shows the estimated marriage prevalence $\psihat(\delta)$ as a proportion; so, $0.3$ corresponds to 30\% of subjects being married in 2010.  The black line is the point estimate of $\psi(\delta)$ ranging across $\delta$; as we discussed previously, because $\psi(\delta)$ is smooth in $\delta$, we can interpolate between the point estimates and plot a smooth line.  The darker blue confidence band is a pointwise 95\% confidence interval that would give us valid inference at a single $\delta$ value.  The lighter blue confidence band is the 95\% confidence interval that allows us to perform inference across the whole range of $\delta$ values.

\medskip

In this example, we see that incarceration has a strong effect on marriage rates: Estimated marriage rates decrease with higher odds of incarceration ($\delta > 1$) but only increase slightly with lower odds of incarceration ($\delta < 1$).  To be more specific, at observational levels of incarceration (i.e., leaving the odds of incarceration unchanged) they estimated a marriage rate of $\psi(1) = 0.294$, or $29.4\%$.  If the odds of treatment were doubled, they estimated marriage rates would decrease to $0.281$, and if they were quadrupled they would decrease even further to $0.236$.  Conversely, if the odds of treatment were halved, they estimated marriage rates would only increase to $0.297$; the estimated marriage rates are the same if the odds are quartered.

\medskip

Finally, we can use the uniform confidence interval in Figure \ref{fig:delta_curve} to test for no effect across the range $\delta \in [0.2, 5]$, and we reject the null hypothesis at the $\alpha = 0.05$ level with a p-value of $0.049$.  By eye, we can roughly see that it is just about impossible to run a horizontal line through the uniform $95\%$ confidence interval.  But it is very close, and the p-value is $0.049$, only just below $0.05$.

%%%%%%%%%%%%%%%%%%%%%%%%%
%%% Extensions 

\section{Extensions \& future directions} \label{sec:extensions}

\subsection{Censoring \& dropout}

In time-varying settings, it is common that subjects are censored or dropout so that we no longer observe them after a certain time point. This complicates causal inferences, in that time-varying covariates, treatment, and outcomes are not observed for all subjects across all time points in the study. The incremental propensity score approach can be extended to account for censoring and dropout (\cite{kim202xincremental}).  As with almost all analyses with censoring, we must assume that the data is ``missing at random'' (MAR) or ``missing completely at random'' (MCAR) if we wish to make progress.  The MCAR assumption is much stronger and thus less likely to be true, so researchers aim to use the weaker MAR assumption.  In our time-varying setup, this corresponds to assuming that, given the past data, subjects' values of future data are independent of whether or not they are censored.  This assumption allows for identification and estimation of the estimand $\psi(\delta)$.  The estimator is very similar to what we outlined in Section \ref{sec:time}, but includes an inverse weighting term for the probability of censoring.  This is common in time-varying causal analyses with missing data (e.g., \cite{robins2000marginal}), and takes the same form as inverse propensity weighting to account for missing potential outcomes.

\subsection{Future directions}

In this chapter, we have outlined the simplest approach to estimate incremental effects without considering any major complications.  As we discussed in the previous section, this approach has been extended to account for time-varying censoring (\cite{kim202xincremental}).  There are many natural extensions that have already been developed for standard estimands that would make sense in the incremental effects paradigm, including continuous treatment effects, conditional treatment effects, instrumental variables analyses, and sensitivity (to unmeasured confounding) analyses.  Additionally, the incremental effects approach generates a few of its own unique problems.  First, as we highlighted in Remarks \ref{remark:del_var_x} and \ref{remark:var_del_t}, we can allow $\delta$ to vary with covariates and time.  Second, we might hope to characterize the stochastic treatment intervention distribution in more detail than just the parameter $\delta$. For example, we may be interested in estimating the average number of treatments, the first treatment, or the modal number of treatments under the incremental propensity score intervention. In this case, we would need to develop an estimator for the distribution of treatment attendance over time for various $\delta$.

%%%%%%%%%%%%%%%%%%%
%%% Discussion

\section{Discussion}
This note serves as an introduction to and a review of \cite{kennedy2019nonparametric}, which describes a class of causal effects based on incremental propensity score interventions. Such effects can be identified even if Positivity is violated and thus can be useful in time-varying settings when the number of possible treatment sequences is large and Positivity is less likely to hold. We have compared this class of effects to other effects commonly estimated in practice, such as average treatment effects or coefficients in marginal structural models, and highlighted the scenarios where it can be most useful. Along the way, we have reviewed the estimation and inferential procedure proposed in \cite{kennedy2019nonparametric} and briefly reviewed the underlying semiparametric efficiency theory.

%%%%%%%%%%%%%%%%%%%%%%
%%% References

\newpage
\section{References}

\bibliographystyle{plainnat}
\bibliography{bibliography}
\end{document}